\documentclass[aps,prb,reprint,showpacs]{revtex4-1}
\usepackage{amsmath,amssymb,mathrsfs}
\usepackage[pdftex]{graphicx}
\usepackage{epstopdf}

\newcommand{\abs}[1]{\left|#1\right|}
\newcommand{\logTerm}[2]{\mathcal{L}_#1\big(#2\big)}

\usepackage{color}

\allowdisplaybreaks[4]

\usepackage{color}

\begin{document}
%%%%%%%%%%%%%%%%%%%%%%%%%%%
\title{Transport properties of a multichannel Kondo dot in a magnetic field}
%%%%%%%%%%%%%%%%%%%%%%%%%%%
\author{Christoph B. M. H\"orig}
\author{Dirk Schuricht}
\affiliation{Institute for Theory of Statistical Physics, 
RWTH Aachen, 52056 Aachen, Germany}
\affiliation{JARA-Fundamentals of Future Information Technology}
\date{\today}
\pagestyle{plain}

\begin{abstract}
We study the nonequilibrium transport through a multichannel Kondo quantum dot in 
the presence of a magnetic field. We use the exact solution of the two-loop 
renormalization group equation to derive analytical results for the $g$ factor, 
the spin relaxation rates, the magnetization, and the differential conductance. 
We show that the finite magnetization leads to a coupling between the conduction 
channels which manifests itself in additional features in the differential conductance. 
\end{abstract}
%05.10.Cc	Renormalization group methods
%05.70.Ln	Nonequilibrium and irreversible thermodynamics
%71.10.-w	Theories and models of many-electron systems
%73.63.-b	 	Electronic transport in nanoscale materials and structures
%73.63.Kv	Quantum dots
\maketitle

%%%%%%%%%%%%%%%%%%%%%%%%%
\emph{Introduction.}---The study of a localized spin coupled via an antiferromagnetic 
exchange interaction $J$ to $K$ independent electronic reservoirs has a long history in 
condensed matter physics.\cite{Hewson93} In the simplest case of $K=1$ the 
electron spins completely screen the local spin at low energies and thus lead 
to a Fermi liquid. In a renormalization group (RG) analysis this situation is 
characterized by the divergence of the renormalized exchange coupling $J(\Lambda)$
at the Kondo scale $\Lambda=T_\text{K}$. The situation is completely 
changed\cite{NozieresBlandin80} if the spin is coupled to more than one 
screening channel ($K>1$). Then the renormalized exchange coupling stays finite and flows
to a non-trivial fixed point\cite{NozieresBlandin80,Gan94} $J^*\sim 1/K$ at low energies,
which manifests itself in unusual non-Fermi liquid behavior like a non-integer ``ground-state 
degeneracy'' or characteristic power laws in various 
observables.\cite{AndreiDestri84prl,AffleckLudwig93}

The recent developments in the ability to engineer devices on the nanoscale lead to 
the experimental realization\cite{Potok-07} of two-channel Kondo physics in a 
quantum dot set-up.\cite{OregGoldhaber-Gordon03} In this set-up it was possible 
to measure the differential conductance and observe universal scaling and  
square-root behavior which are characteristic for the two-channel Kondo effect. This
triggered theoretical studies\cite{2CK,MitraRosch11,Eflow} of the transport properties of 
multichannel systems using conformal field theory as well as numerical and perturbative
RG methods. The latter uses a perturbative expansion in the renormalized exchange
coupling which is well-controlled provided $K\gg 1$. Specifically, 
Mitra and Rosch\cite{MitraRosch11} calculated the differential conductance, the splitting of the 
Kondo resonance in the $T$ matrix, and the current-induced decoherence in the absence 
of a magnetic field. Recently the spin dynamics was studied\cite{Eflow} in the absence of a 
bias voltage and shown to possess pure power-law decay with an exponent $g=4/K$.

In this Brief Report we extend the analysis of Mitra and Rosch\cite{MitraRosch11}
to include an external magnetic field. We perform 
a real-time RG (RTRG) analysis\cite{Schoeller09} to derive the renormalized magnetic
field, the spin relaxation rates, the magnetization of the quantum dot, and the current. 
In particular, we focus on inelastic cotunneling processes which lead to characteristic features
in the differential conductance whenever one of the applied bias voltages $V_i$'s equals the 
value of the renormalized magnetic field. We further show that the finite magnetization
on the dot leads to a coupling of the conduction channels which results in additional 
features in the differential conductance.

%%%%%%%%%%%%%%%%%%%%%%%%%
\begin{figure}[b]
	\centering
	\includegraphics[width=0.35\textwidth]{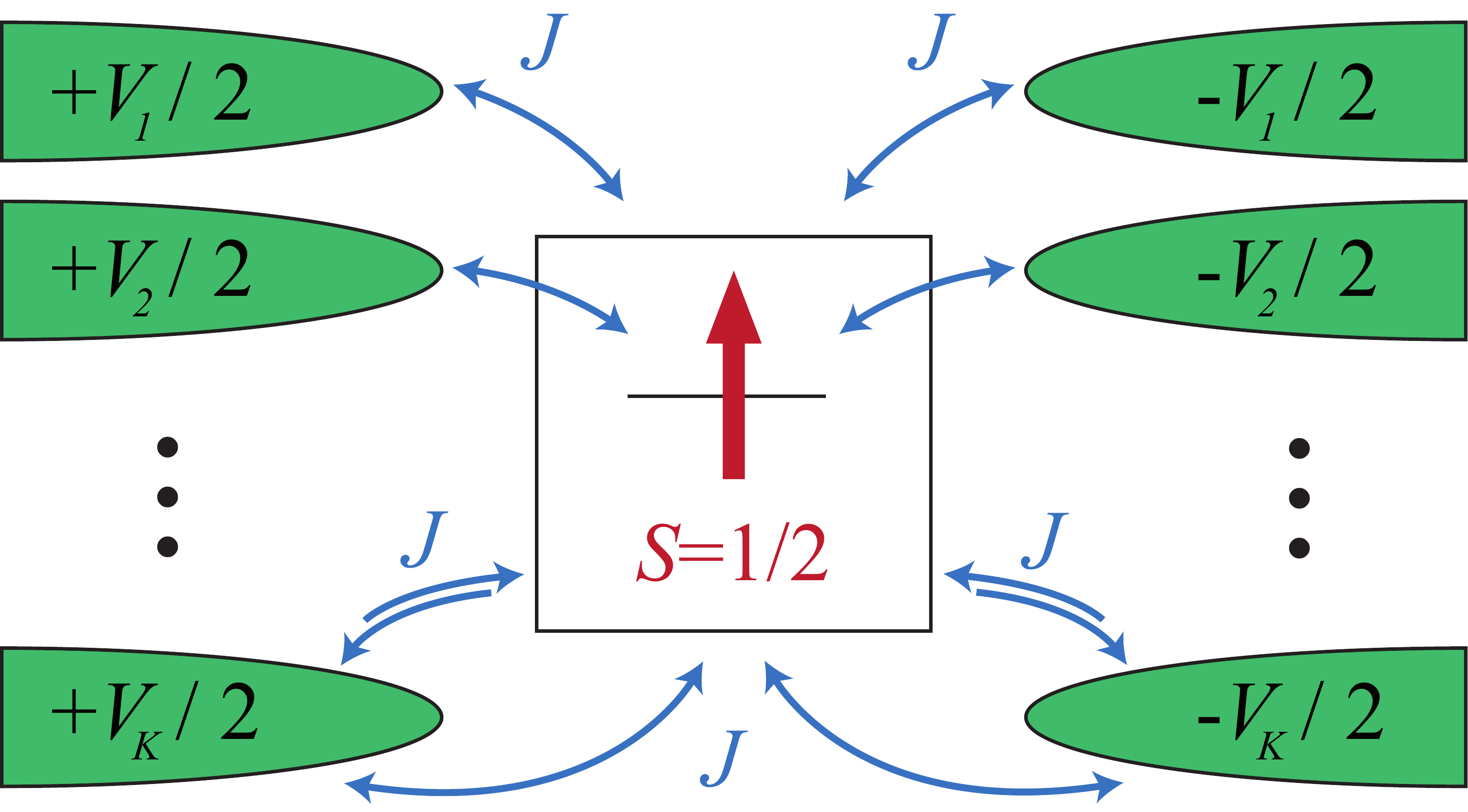}
	\caption{\label{fig:model}(Color online) Sketch of the $K$-channel Kondo model 
	\eqref{eq:ham}. The channels are labeled by $i=1,\ldots,K$, each channel is 
	decomposed into two leads $\alpha=L,R$ held at chemical potentials 
	$\mu_{L/R}^i=\pm V_i/2$. The spin-1/2 on the dot is subject to a magnetic field 
	$h_0$. The exchange interaction $J$ between dot and leads present in all 
	channels is depicted for the $K$th channel explicitly.}
\end{figure}
\emph{Model.}---We consider a quantum dot possessing a spin-1/2 degree of freedom, 
which is coupled via an exchange interaction $J$ to $K$ independent electronic reservoirs. 
At low energies each reservoir constitutes one screening channel for the local spin
thus leading to an over-screened situation for $K>1$. We will thus use the terms reservoir and 
channel interchangeably.
Furthermore, the spin-1/2 is subject to an external magnetic field $h_0$. 
Each reservoir consists  of a left ($L$) and right ($R$) lead which are held at chemical 
potentials $\mu_\alpha^i$, $\alpha=L,R$, $i=1,\ldots,K$ (see Fig.~\ref{fig:model}). 
Specifically we consider the system
\begin{equation}
  \begin{split}
		H=&\sum_{i\alpha k\sigma}\epsilon_k\,c_{i\alpha k\sigma}^\dagger 
		c_{i\alpha k\sigma}+ h_0\,S^z \\
		&+\frac{J_0}{2\nu_0} \sum_{i\alpha\alpha' k k' \sigma\sigma'}\vec{S}
		\cdot\vec{\sigma}_{\sigma\sigma'}
		c_{i\alpha k\sigma}^\dagger c_{i\alpha' k'\sigma'}.
  \end{split}
  \label{eq:ham}
\end{equation}
Here $c_{i\alpha k\sigma}^\dagger$ and $c_{i\alpha k\sigma}$ create and annihilate 
electrons with momentum $k$ and spin $\sigma=\uparrow,\downarrow$ in lead $\alpha$
of the reservoir $i$, $\vec\sigma$ denotes the Pauli matrices, and $\vec{S}$ is the spin-1/2
operator on the dot. The antiferromagnetic exchange coupling $J_0>0$ is 
dimensionless in our convention. We stress that the exchange term does not couple 
different reservoirs. The chemical 
potentials in the leads are parametrized by $\mu_{L/R}^i=\pm V_i/2$ thus applying a 
bias voltage $V_i$ to each reservoir. Furthermore we introduce an ultra-violet cutoff $D$ 
in each reservoir via the density of states $N(\omega)=\nu_0D^2/(D^2+\omega^2)$.
In the absence of a magnetic 
field the transport properties of the model \eqref{eq:ham} have been previously studied 
in Ref.~\onlinecite{MitraRosch11}.

%%%%%%%%%%%%%%%%%%%%%%%%%
\emph{RG analysis.}---Following Ref.~\onlinecite{Schoeller09} we have performed 
a two-loop RTRG analysis including a consistent derivation of the relevant 
relaxation rates. This method has been successfully applied to study transport 
properties of other Kondo-type quantum dots in the past.\cite{Schoeller09,PS11} 
The starting point is an RG equation for the renormalized exchange coupling 
$J(\Lambda)$ obtained by integrating out the high-energy degrees of freedom in 
the reservoirs. To accomplish this the one introduces a cutoff $\Lambda$ into the
Fermi function and integrates out the Matsubara poles on the imaginary axis by 
decreasing the cutoff from its initial value $\Lambda_0\sim D$ down to some physical 
energy scale. The RG equation for the $K$-channel model 
\eqref{eq:ham} reads up to two loop
\begin{equation}
	\Lambda\frac{d}{d\Lambda}J=\beta(J)=-2J^2(1-KJ),
	\label{eq:PMSeq}
\end{equation}
which defines the reference solution $J(\Lambda)$ for our analysis.
The RG flow has the well-known\cite{NozieresBlandin80} non-trivial fixed point $J^*=1/K$. 
The scaling dimension of the leading irrelevant operator is $\Delta=\beta'(J^*)=2/K$,
which is valid for $K\gg 1$ while the exact result is given by\cite{AffleckLudwig93} 
$\Delta=2/(K+2)$. The RG equation possesses the invariant
\begin{equation}
	T_\text{K}=\Lambda_0\left(\frac{eJ_0}{J^*-J_0}\right)^{K/2}e^{-1/2J_0},
\end{equation}
which defines the Kondo temperature. With the initial condition $J_0=J(\Lambda_0)$ 
the solution of the RG equation can be explicitly written as
\begin{equation}
J(\Lambda)=\frac{J^*}{1+W(z)},\quad
	z=\left(\frac{\Lambda}{T_\text{K}}\right)^\Delta.
	\label{eq:coupling}
\end{equation}
Here $W(z)$ denotes the Lambert W function\cite{Corless96} defined by 
$z=W(z)e^{W(z)}$, which satisfies $W'(z)={W(z)/z/(1+W(z))}$ for $z\neq 0$. The fixed 
point $J^*$ is reached for $\Lambda\to 0$ as $W(0)=0$. We note that the solution
\eqref{eq:coupling} is valid for all $J_0$ but, of course, the derivation of \eqref{eq:PMSeq}
requires $J_0\ll 1$. If $J_0<J^*$ we can perform the scaling limit $\Lambda_0\to\infty$
and $J_0\to 0$ while keeping the Kondo temperature constant. In this limit the solution 
for $\Lambda\ll T_\text{K}$ simplifies to
\begin{equation}
	J(\Lambda)=J^*\left[1-\left(\frac{\Lambda}{T_\text{K}}\right)^\Delta\right],
	\label{eq:coupling_SL}
\end{equation}
\emph{i.e.} there is a characteristic power-law behavior. Observables are calculated in a 
systematic expansion around the reference solution and can be expressed
in terms of the renormalized exchange coupling $J_c\equiv J(\Lambda_c)$ at the physical 
energy scale\cite{scale} (we consider $T=0$)
\begin{equation}
\Lambda_c=\max\{V,h_0\}.
\end{equation}
In contrast to the  one-channel Kondo model the existence of the attractive fixed
point $J^*=1/K$ implies that this expansion is well-defined for all $\Lambda_c$ provided 
$K\gg 1$. We have calculated the effective dot Liouvillian and the current kernel 
yielding the renormalized magnetic field, the spin relaxation rates, the dot magnetization, 
and the current including the leading logarithmic corrections. All calculations follow 
Ref.~\onlinecite{Schoeller09}, we present here only the results and discuss their
properties.

%%%%%%%%%%%%%%%%%%%%%%%%%
\begin{figure}[t]
	\centering
	\includegraphics[width=0.49\textwidth]{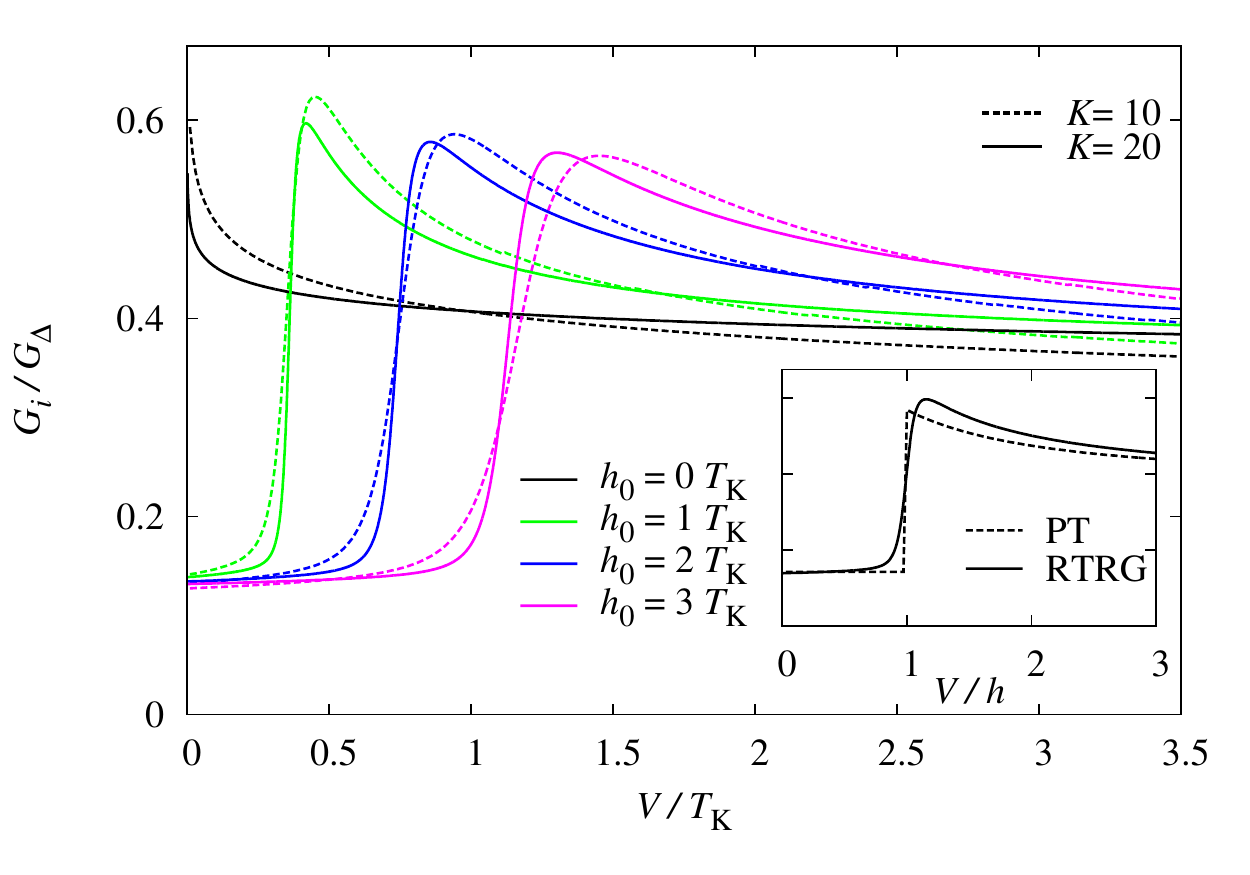}
	\caption{\label{fig:conductance-identical}(Color online) Differential conductance 
	of an individual channel for identical bias voltages $V_i=V$ and $K=10$ (dashed lines) 
	or ${K=20}$ (solid lines) channels. We use the normalization 
	$G_\Delta=(\frac{2e^2}{h})\frac{3\pi^2}{16}\Delta^2$. For zero field one 
	finds\cite{MitraRosch11} $G/G_\Delta=1-2(V/T_\text{K})^\Delta$, \emph{i.e.} 
	$G/G_\Delta\to 1$ for $V\to 0$.
	Inset: Comparison of RTRG (solid line) and renormalized perturbation theory (PT) up to 
	$\mathcal{O}(J_c^2)$ (dashed line) for $h_0=T_\text{K}$ and $K=20$. We note that
	the broadening of the feature at $V=h$ is not captured in PT.}
\end{figure}
\emph{Identical bias voltages.}---First let us assume that all bias voltages are identical,
\emph{i.e.} $V_i=V$. Straightforward calculation yields the renormalized magnetic field
\begin{equation}
	h=\bigl[1- K (J_c-J_0)\bigr]h_0, \label{eq:h}
\end{equation}
which gives for the $g$ factor in leading order (we consider the scaling limit $J_0\to 0$
from now)
\begin{equation}
	g=2\frac{\partial h}{\partial h_0}
	=2(1-K J_c)
	\stackrel{\Lambda_c\ll T_\text{K}}= 2\left(\frac{\Lambda_c}{T_\text{K}}\right)^\Delta.
\end{equation}
The longitudinal and transverse spin relaxation rates 
\begin{eqnarray}
	\Gamma_1 &=&\pi
	\left[h + \frac12 \big(\abs{V - h}_2+ V + h\big)\right] K J_c^2,\label{eq:gamma1}\\
	\Gamma_2 &=&\frac{\pi}{2}\left[V + h + \frac12 \big(\abs{V - h}_1 + V + h\big)\right] 
	K J_c^2,\label{eq:gamma2}
\end{eqnarray}
with 
\begin{equation}
	\abs{x}_l=\frac{2}{\pi}x\arctan\frac{x}{\Gamma_l},\quad l=1,2.
\end{equation}
We note that for $\Lambda_c\gg T_\text{K}$ the renormalization of the magnetic field 
\eqref{eq:h} is much stronger than in the one-channel Kondo model, as the spin on the dot 
is coupled to more screening channels.

Explicit formulas for the dot magnetization and the current are given in Eq.~\eqref{eq:MI} 
below for the case of different bias voltages. Simplifying to $V_i=V$ we obtain the 
differential conductance $G_i=dI_i/dV_i$ per channel, which is plotted in 
Fig.~\ref{fig:conductance-identical}. The conductance sharply increases around
$V=h$ where inelastic cotunneling processes start to contribute. Due to the strong 
renormalization of the magnetic field the increase occurs at voltages much smaller than the
applied magnetic field $h_0$. Close to the resonance we find
\begin{equation}
	G_i=
	\begin{cases}
		\frac{\pi}{4}J_c^2\bigl[1+2 J_c\logTerm{2}{V-h}\bigr]&\text{for }V<h,\\[2mm]
		\pi J_c^2\bigl[1+2\pi J_c\logTerm{2}{V-h}\bigr]&\text{for }V>h,
	\end{cases}
	\label{eq:conductance}
\end{equation}
where we assumed $|V-h|\ll h$ and defined 
\begin{equation}
	\logTerm{l}{x}=\ln\frac{\Lambda_c}{\sqrt{x^2+\Gamma_l^2}},\quad l=1,2.
\end{equation}
We note that \eqref{eq:conductance} is formally identical to the differential conductance 
in the one-channel Kondo model\cite{Schoeller09} except for the functional form of the 
renormalized coupling \eqref{eq:coupling_SL}. In particular, there is no explicit dependence 
on the number of channels. We further note that close to the fixed point 
the conductance \eqref{eq:conductance} is a quantity of order $1/K^2$.
The logarithmic divergencies at $V=h$ are cut off by the transverse spin relaxation rate
$\Gamma_2$. Thus, as $\Gamma_2\sim 1/K$, this feature becomes sharper with increasing
$K$. At small voltages $V<h$ the conductance is independently of the bias voltage 
given by $G_i=\pi J_c^2/4$. In particular, a power-law behavior in the voltage is only 
found for vanishing magnetic field\cite{MitraRosch11} while \eqref{eq:coupling_SL}
yields a power-law dependence on the magnetic field. In the limit of a large field 
$h\gg T_\text{K}$ the linear conductance can be derived using\cite{Corless96} 
$W(z)\sim\ln z$ for $z\to\infty$ to be $G_i(V=0,h)=\pi/16/\ln^2(h/T_\text{K})$, 
which is identical to the result in the one-channel model.

%%%%%%%%%%%%%%%%%%%%%%%%%
\begin{figure}[t]
	\centering
	\includegraphics[width=0.49\textwidth]{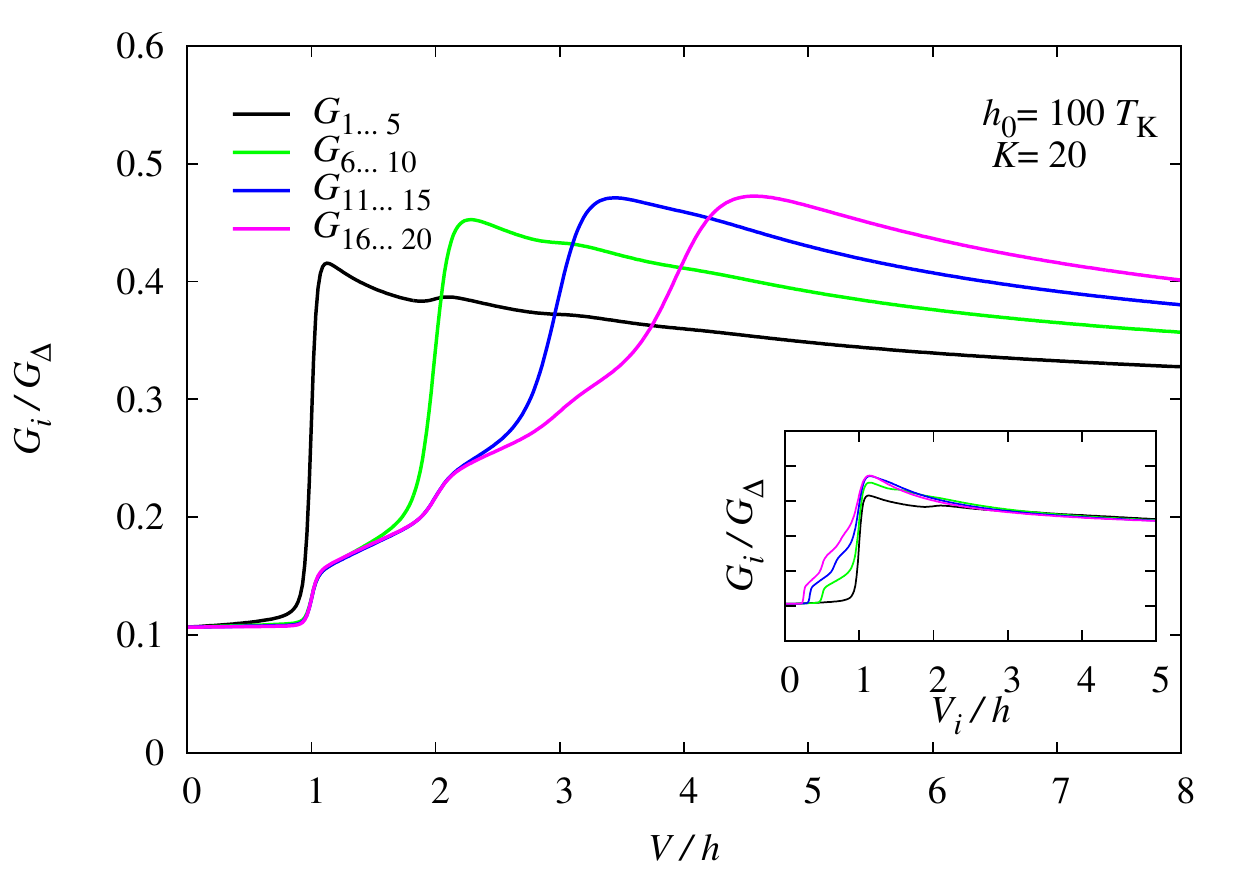}
	\caption{\label{fig:conductance-different}(Color online) 
	Differential conductance $G_i$ through channel $i$ for $h_0=100\,T_K$ and $K=20$. 
	The bias voltages are applied such that series of five channels have identical 
	voltages, \emph{i.e.} $V_i=a_i V$ with $a_{1,\ldots,5}=1$, $a_{6,\ldots,10}=1/2$, 
	$a_{11,\ldots,15}=1/3$, and $a_{16,\ldots,20}=1/4$. For $h_0=0$ the conductance 
	is similar to the black curve in Fig.~\ref{fig:conductance-identical}, \emph{i.e.}
	there are no features.
	Inset: Same situation as in the main panel, but plotted with respect 
	to the respective bias voltage $V_i$.}
\end{figure}
\emph{Different bias voltages.}---In the following we relax the condition $V_i=V$ and 
consider channel-dependent bias voltages $V_i$. This introduces further energy 
scales which will give rise to additional features in the dot magnetization and thus the 
differential conductance. In this general set-up the magnetization
and current through the $i$th channel are given by
\begin{equation}
	M=-\frac{f_1}{2f_2}, \quad I_i=f_I^i+2Mf_M^i
	\label{eq:MI}
\end{equation}
where $f_{1,2}$ denote rates appearing in the Liouvillian and $f^i_{I,M}$ are similar
terms in the current kernel. Explicitly these rates read
\begin{eqnarray}
	f_1
		&=& 
			2\pi K J_c^2 h + 4\pi K J_c^3 h\logTerm{2}{h}\nonumber\\*
		&&	-2\pi J_c^3\sum\nolimits_{j=1\ldots K}(V_j-h)\logTerm{2}{V_j-h},\label{eq:f1}\\
	f_2
		&=&
			\pi K J_c^2 h+ 2\pi K J_c^3 h\logTerm{2}{h}\nonumber\\\*
			&&+\frac\pi2J_c^2
			\sum\nolimits_{j=1\ldots K}\big(\abs{V_j-h}_2+V_j+h\big)\label{eq:f2}\\*
			&&+\pi J_c^3\sum\nolimits_{j=1\ldots K}
			\big(\abs{V_j-h}_2-V_j+h\big)\logTerm{2}{V_j-h},\nonumber\\
	f_I^i
		&=&
			\frac34\pi J_c^2 V_i+\pi J_c^3 V_i\logTerm{1}{V_i}\nonumber\\*
			&&+\pi J_c^3(V_i-h)\logTerm{2}{V_i-h},\label{eq:f3}\\
	f_M^i
		&=&
			-\frac\pi4 J_c^2\big(\abs{V_i-h}_2-V_i-h\big)+
			\pi J_c^3V_i\logTerm{1}{V_i}\nonumber\\*
			&&+\pi J_c^3 h\logTerm{2}{h}
			-\frac\pi2 J_c^3\abs{V_i-h}_2\logTerm{2}{V_i-h}.\label{eq:f4}
\end{eqnarray}
The relaxation rates for the case of different bias
voltages are obtained by straightforward generalization of 
Eqs.~\eqref{eq:gamma1} and~\eqref{eq:gamma2}. 
In the derivation of Eqs.~\eqref{eq:f1}--\eqref{eq:f4} 
we have neglected all terms in order $J_c^3$ that 
do not contain logarithms at either $h=0$, $V=0$, or $V=h$. Thus when calculating the 
magnetization one has to expand consistently up to this order.

\begin{figure}[t]
	\centering
	\includegraphics[width=0.49\textwidth]{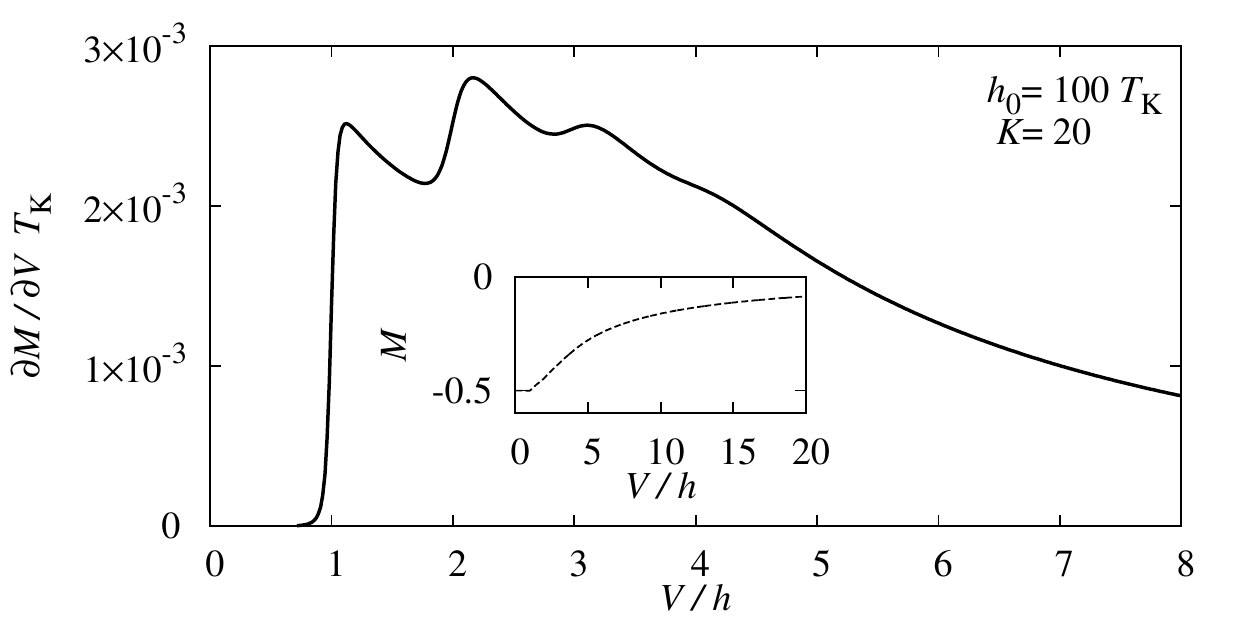}
	\caption{\label{fig:magnetization-different}
	Derivative of the magnetization with respect to the voltage, 
	\emph{i.e.} $\partial M/\partial V$. All parameters are as in 
	Fig.~\ref{fig:conductance-different}. We observe pronounced 
	features at $V=h$ and $V=2h$. 
	Inset: Magnetization as a function of the bias voltage.}
\end{figure}
We stress that although the different channels are not directly coupled via exchange 
interactions, the finite magnetization $M$ mediates a feedback between them. Consider 
for example the differential conductance of channel 1, \emph{i.e.} the black curve in 
Fig.~\ref{fig:conductance-different}. The sharp increase at $V=V_1=h$ is again due to 
the onset of inelastic cotunneling processes. However, there is a second feature 
at $V\approx 2h$ which is caused by the non-trivial voltage dependence of the dot 
magnetization. This can be seen in the derivative $\partial M/\partial V$ shown in 
Fig.~\ref{fig:magnetization-different}, which directly enters the differential conductance
[see \eqref{eq:MI}]. Similarly, the conductance in channel 6 (green curve in 
Fig.~\ref{fig:conductance-different}) possesses a feature at $V=2V_6=h$ which is caused by 
the effect of the applied voltage in channels 1 to 5 onto $M$, while the increase at 
$V_6=h$ (\emph{i.e.} $V=2h$) is due to the onset of inelastic cotunneling in channel 6.
In this way the nonequilibrium magnetization introduces additional features into 
the individual conductances. We stress that such coupling effects between the
channels are absent for vanishing magnetic field\cite{MitraRosch11} or if all applied 
bias voltages are identical (see Fig.~\ref{fig:conductance-identical}). 

As a special case of the general set-up with channel-dependent bias voltages we 
can recover the experimental situation realized\cite{Potok-07} by Potok~\emph{et al.} 
in a semiconductor quantum dot. This is achieved by setting the chemical potentials 
to $\mu_{L/R}^1=\pm V/2$ and $\mu_{L/R}^i=0$ for $i=2,\ldots,K$. In each of the 
channels $2,\ldots,K$ we introduce even and odd combinations of the 
electron operators, $(c_{iLk\sigma}\pm c_{iRk\sigma})/\sqrt{2}$, such that the 
even combinations couple to the spin on the dot with exchange interaction $2J$ 
while the odd ones decouple completely. The resulting RG equation for $J$ is given 
by \eqref{eq:PMSeq} and thus possesses the fixed point $J^*$. 
The experimental set-up~\cite{Potok-07} is now obtained by specializing to $K=2$.
In the presence of a magnetic field the resulting differential conductance is very 
similar to the case of identical bias voltages shown in Fig.~\ref{fig:conductance-identical}.
In particular, since the second channel does not provide an additional energy scale
there appear no features in the differential conductance beside the cotunneling peak
at $V=h$. A possible experimental set-up to observe the additional features shown 
in Fig.~\ref{fig:conductance-different} thus requires at least two channels with 
non-zero and different bias voltages.

%%%%%%%%%%%%%%%%%%%%%%%%%
\emph{Conclusions.}---To sum up, we have studied the nonequilibrium transport properties 
of a multichannel Kondo quantum dot in the presence of a magnetic field. We used 
the solution of the two-loop RG equation to derive analytical results for the 
$g$ factor, the spin relaxation rates, the dot magnetization, and the 
differential conductance. The latter shows typical features of inelastic cotunneling. 
We showed that the main difference to the previously studied\cite{MitraRosch11} situation 
without magnetic field is the appearance of additional features in the 
differential conductance, which originate in the feedback between the
channels mediated by the finite dot magnetization.

We thank S. Andergassen, A. Rosch, and H. Schoeller for useful discussions.
This work was supported by the German Research Foundation (DFG) through the 
Emmy-Noether Program.


\begin{thebibliography}{10}
\bibitem{scale}
For the numerical evaluation we use ${\Lambda_c=\sqrt{V^2+h_0^2}}$.
\end{thebibliography}
\end{document}